\documentclass[useAMS,usenatbib,usegraphicx]{mn2e}

\def\km{K$'$}
\def\ks{K$_s$}
\def\chisq{$\chi^2$}
\def\micron{$\mu$m}
\def\h2o{H$_{\rm 2}$O}
\def\co2{CO$_{\rm 2}$}
\def\o3{O$_{\rm 3}$}
\def\teff{${\rm T}_{\rm eff}$}

\title{Limits on the infrared photometric monitoring of brown dwarfs} 

\author[C.A.L.\ Bailer-Jones and M.\
Lamm]{C.A.L.\ Bailer-Jones\thanks{Email: calj@mpia-hd.mpg.de} and M.\
Lamm\\ Max-Planck-Institut f\"ur Astronomie, K\"onigstuhl 17, D-69117
Heidelberg, Germany}

\begin{document}

\date{Submitted 7 August 2002; Accepted 17 October 2002}

\maketitle

\label{firstpage}

\begin{abstract}
Recent monitoring programs of ultra cool field M and L dwarfs (low
mass stars or brown dwarfs) have uncovered low amplitude photometric
I-band variations which may be associated with an inhomogeneous
distribution of photospheric condensates.  Further evidence hints that
this distribution may evolve on very short timescales, specifically of
order a rotation period or less.  In an attempt to study this
behaviour in more detail, we have carried out a pilot program to
monitor three L dwarfs in the near infrared where these objects are
significantly brighter than at shorter wavelengths.  We present a
robust data analysis method for improving the precision and
reliability of infrared photometry.  No significant variability was
detected in either the J or \km\ bands in 2M1439 and SDSS1203 above a
peak-to-peak amplitude of 0.04 mag (0.08 mag for 2M1112). The main
limiting factor in achieving lower detection limits is suspected to be
second order extinction effects in the Earth's atmosphere, on account
of the very different colours of the target and reference stars.
Suggestions are given for overcoming such effects which should improve
the sensitivity and reliability of infrared variability searches.

\end{abstract}


\section{Introduction}\label{introduction}

Thermodynamical considerations show that dust particles form in the
atmospheres of ultra cool M, L and T dwarfs (e.g.\ Allard et al.\
\cite{allard01}; Burrows \& Sharp \cite{burrows99}; Helling
\cite{helling01}; Lodders \cite{lodders99}; Lunine et al.\
\cite{lunine89}; Sharp \& Huebner \cite{sharp90}; Tsuji et al.\
\cite{tsuji96}).  This has been confirmed in many cases from the
modelling of their optical and near infrared spectral energy
distributions (e.g.\ Leggett
et al.\ \cite{leggett01}; Marley et al.\ \cite{marley02}; Schweitzer
et al. \cite{schweitzer01}\cite{schweitzer02}). Physical models
suggest that as the effective temperature drops below about 2500\,K,
the photosphere becomes increasingly dusty due to increased gas
condensation, whereas at even lower temperatures the dust opacity is
reduced due to gravitational settling of the dust below the
photosphere (Ackerman \& Marley \cite{ackerman01}; Tsuji
\cite{tsuji01}).

However, such models assume uniform horizontal distributions of this
dust, whereas recent observational work has shown a significant
fraction of ultra cool dwarfs to be photometrically variable at the
level of 0.01--0.08 mag (Bailer-Jones \& Mundt
\cite{bjm99}\cite{bjm01}; Clarke et al.\
\cite{clarke02a}\cite{clarke02b}; Gelino et al.\ \cite{gelino02};
Mart\'\i n et al.\ \cite{martin01}; Nakajima et al.\
\cite{nakajima00}; Tinney \& Tolley \cite{tinney99}).  Bailer-Jones \&
Mundt \cite{bjm01} found evidence for a rapid evolution of features on
the surfaces of a few field L dwarfs. Being field objects, they are
presumably part of an older population which have long lost their
disks (age\,$>10$\,Myr), thus ruling out accretion hot spots or other
star--disk activity as a cause of the variability.  This leaves
magnetic star spots or dust clouds as the most plausible candidates,
although as yet there is no {\it direct} evidence for the physical
nature of these surface features.  Theoretical arguments from Gelino
et al.\ \cite{gelino02} and Subhanjoy et al.\ \cite{subhanjoy02} argue
against the existence of spots in ultra cool dwarfs due to the low
ionization fractions.  On the other hand, a non-monotonic variation
with spectral type in the strength of the FeH band at 9896\,\AA\
(which would be effected by iron condensation) noted by Burgasser et
al.\ \cite{burgasser02} may be indicative of an inhomogeneous dust
cloud coverage, at least around the L/T transition.

If the variability is due to dust, then understanding its physical,
chemical and dynamical nature is central to determining the
fundamental physical parameters of ultra cool dwarfs (\teff,
abundances, ages etc.) and hence their formation mechanisms.  Using
the dusty and condensed dust models of Allard et al.\ \cite{allard01},
one of us has recently predicted the observational signatures of
different dust cloud and spot scenarios (Bailer-Jones
\cite{bj02}). This shows that there would be a distinct signature in
the J and K bands for these phenomena,
depending on effective temperature and feature size.

In this paper, we present the results of a pilot project to look for
variability in ultra cool dwarfs through near-simultaneous J and \km\
band monitoring. A comparison of the amplitudes and correlation of any
variability can be compared with surface feature scenarios from
various models.

\section{Data acquisition and reduction}

\begin{table*}
\begin{minipage}{125mm}
\caption[]{Targets observed. The reference is the discovery paper and
provides the spectral type and IAU name.  The photometry is from 2MASS
from the archive compiled by Kirkpatrick \cite{kirkpatrick03}. }
\label{targets}
\begin{tabular}{lllrrrl}
\hline
name 		& IAU name			& SpT	& J	& \ks\	& reference	\\
\hline
2M1439		& 2MASSW J1439284$+$192915	& L1	& 12.76	& 11.58	& Reid et al.\ \cite{reid00}	\\
SDSS1203	& SDSSp  J120358.19$+$001550.33	& L3	& 14.02 & 12.48	& Fan et al.\  \cite{fan00}      \\
2M1112		& 2MASSW J1112257$+$354813	& L4.5	& 14.57	& 12.69	& Kirkpatrick et al.\ \cite{kirkpatrick00} \\
\hline
\end{tabular}
\end{minipage}
\end{table*}

Details of the three targets monitored are given in
Table~\ref{targets}. They were chosen from a list of sufficiently
bright objects with appropriate RA/Dec for the observatory/season to
represent a range of L spectral types.  The targets were observed over
16 nights in March/April 2001 from the 1.23m telescope at Calar Alto,
Spain, with the MAGIC camera, equipped with a 256$\times$256 NICMOS3
array. The pixel scale was 1.15$''$/pix providing a field-of-view of
5$'\times$5$'$.  The objects were observed in both the J (50\%
transmission points at 1.12 and 1.40\micron) and \km\ (50\%
transmission points at 1.93 and 2.27\micron) filters.  Each object was
monitored continuously for several hours each night (alternating
between J and \km), weather conditions permitting. Table~\ref{obslog}
shows the log of observations.  The seeing varied between about
1.2$''$ and 2.2$''$ in J (poorer seeing images are not retained in the
final light curves). Weather conditions varied across the run.

\begin{table*}
\begin{minipage}{112mm}
\caption[]{Log of number of observations retained in final light curve
for each object. AJD is defined as JD$-$2450000.}
\label{obslog}
\begin{tabular}{rlrrrrrrl}\hline
Night & AJD & \multicolumn{2}{r}{2M1112} & \multicolumn{2}{r}{SDSS1203} & \multicolumn{2}{r}{2M1439} & weather \\
      &     	& J	& \km\	& J	& \km\	& J	& \km\	& \\
\hline
~2    & 1990.5  & 24	& 27	&	&	& 	&	& some cloud \\
~3    & 1991.5  &  8	& 10	&	&	& 	&	& cloudy \\
~4    & 1992.5  & 12	& 13	&	&	& 41	& 41	& mostly clear \\
~5    & 1993.5  & 11	& 12	&	&	& 21	& 21	& clear \\
~6    & 1994.5  &	&	&	&	& 32	& 32	& some cloud, later clear \\
~7    & 1995.5  &  3	&  3	&	&	& 42	& 42	& some cloud, later clear \\
~8    & 1996.5  &	&	&	&	& 	& 	& rain (no observations) \\
~9    & 1997.5  &	&	&	&	& 6 	& 6	& high humidity and cloud \\
10    & 1998.5  &	&	& 7	& 7	& 26	& 26	& mostly clear \\
11    & 1999.5  &	&	&	&	& 	&	& fog (no observations) \\
12    & 2000.5  &	&	&	&	& 	&	& fog (no observations) \\
13    & 2001.5  &	&	& 39	& 39	& 10	& 10	& clear \\
14    & 2002.5  &	&	& 42	& 42	& 10	& 10	& clear \\
15    & 2003.5  & 43	& 44	&	&	& 6	& 6	& clear \\
16    & 2004.5  &	&	&	&	& 15	& 15	& clear, high humidity \\
17    & 2005.5  &	&	& 16	& 16	& 23	& 23	& mostly clear \\
\hline
Total &	        & 101	& 109	& 104	& 104	& 232	& 232	\\
\hline
\end{tabular}
\end{minipage}
\end{table*}

A problem with the NICMOS arrays is the variation of the quantum
efficiency (QE) across a single pixel (intra-pixel QE variation). On
account of the coarse pixel scale and undersampled photometry, this
could result in a variation in the detected flux of a non-variable
star as its image moves relative to the center of a pixel. For this
reason, each {\it epoch} (point in the final light curves) is
derived from 45 frames taken with a dithering macro with non-integer
pixel offsets:  A field is observed at each of the nine points in a
3$\times$3 square grid, with x/y distance 12$''$ (10.4 pixels) between
grid points. Five consecutive 5s integrations ({\em repeats}) are
taken at each grid point, read out in double correlated readout (DCR)
mode. The macro required 4.7 minutes to execute.  Due to a variable
bias level, the first repeat at each grid point could not be
used, leaving a total of 36 usable frames. A clipped average of these
points provides a single point in the final light curve (see
section~\ref{photometry}).  This procedure is necessary to achieve high
precision photometry with infrared arrays.

It was further found that the bias levels for repeats 2--5 were all
different, although stable in time. Thus the flat fielding and sky
subtraction had to be performed separately for each repeat.

Flat fields were created on each night from the differences between
illuminated and non-illuminated dome flats.  For each epoch, a sky
image was then created from a median combination of the nine
frame positions for that repeat stage (each offset to have a common flux zero
point to accommodate changes in the overall background). This process
removed all sources, leaving an image of the 2D background
(predominantly sky brightness) variations which was subtracted from
each frame.

No correction was made for nonlinearity in the flux response of the
pixels. In principle, a nonlinearity can introduce apparent
variability if the stars do not remain exactly in the same place on
the detector or if the stars or background undergo a uniform change in
brightness (e.g.\ due to changes in sky brightness or atmospheric
extinction). However, we have calculated that such effects introduce
errors in the relative photometry of much less than 1\% under good
conditions, and no more than 1\% in the very worst case.  Note that
bright reference stars were avoided to reduce the influence of
nonlinearity.

\section{Photometry}\label{photometry}

A reference image was established and the positions of the target and
reference stars in this propagated to all other images of the
field. This avoids shifting and interpolating images, which can only
degrade the data quality. Photometry was carried out using the CCDCAP
aperture photometry program written by K.\ Mighell. CCDCAP was
developed to do accurate aperture photometry with small, undersampled
apertures, using a bilinear pixel interpolation algorithm (see
appendix B of Mighell \& Rich \cite{mighell95}). After detailed
investigation of the magnitude error (see below) as a function of
aperture size, aperture radii of 2.4 and 2.0 pixels were used for J
and \km\ respectively. A `hardness' of 1.0 was used in CCDCAP, which
corresponds to the maximum subdivision of a pixel. A `drift' of 1.5
pixels in CCDCAP allowed for optimal recentering of the aperture.

Variability monitoring requires a reliable assessment of the
photometric errors. Error determinations based on readout noise and
Poisson statistics in the source and sky generally underestimate the
true error because they ignore various systematic errors. The grid
observing method outlined above is ideally suited for determining an
empirical error measure based on the 36 frames taken in rapid
succession. However, due to numerous bad pixels -- up to four of the
nine grid positions could be contaminated -- simply taking the mean
and standard error of these measures frequently gave poor results,
because the bad pixels could bias the photometry by several
magnitudes.  As the PSF is undersampled, interpolation over bad pixels
(and cosmic rays) was not deemed reliable (and this can only be done
to make reduction convenient: it does not add information).  Instead,
an iterative rejection scheme was developed to remove bad photometry.
First the median and its standard deviation, $\sigma$, were determined
from all 36 measures. If $\sigma$ exceeded a threshold, the highest 8
and lowest 8 measures were rejected, typically corresponding to all
measures at four grid positions. The median is then
re-established. Iterative clipping of points more than n-$\sigma$
about the median was then done to remove outliers (each iteration
starting with all 36 measures so that re-inclusion was possible). The
initial threshold stage is required because the median of 36 points
containing just a few bad measures can result in an unrepresentative
median and hence large amounts of data being clipped. A threshold of
0.2 mag and n$=$5 were used, although the results are not sensitive to
these values. The iterative clipping converged after 0--3
iterations. This procedure was found to be very robust, both excluding
obvious bad pixels but resulting in more than 32 (of 36) measures
being retained in the majority of cases.

The mean of the remaining $N$ measurements was taken to be the final
magnitude for that epoch. As these are all independent measurements of
the same thing,\footnote{This assumes that the star does not vary over
4.7 minutes. If it does, then this error is an overestimate of the
true error.} the appropriate error, $\delta m$, in this measure is the
standard error in this mean, i.e.\ $\sigma/\sqrt{N}$.  We found that
the theoretical errors produced by IRAF's apphot task (based on source
Poisson noise and detector read-out noise) --  using the same aperture
sizes -- were as much as ten times smaller, i.e.\ grossly
underestimated. This demonstrates the importance of using empirical
error measures in infrared photometry, at least with NICMOS3 arrays.

Photometry was obtained in this manner for the target and reference
stars in all frames. Differential photometry was performed as
described in Bailer-Jones \& Mundt \cite{bjm01} to produce a light
curve for the target relative to the average flux of the reference
stars.  This removes first order (wavelength independent)
variations in the sky extinction between epochs.\footnote{This
is true provided the extinction is constant across the 5$'$ image
field size averaged across the $70 \cos \delta$ arcmins on the sky
through which the field moves during the 4.7 min macro execution
time.}  Light curves were similarly constructed for each reference star
relative to all the other reference stars. 6, 8 and 5 reference stars
were retained for 2M1112, SDSS1203 and 2M1439 respectively.

In what follows, {\it light curve} shall always mean the relative
magnitude light curve, i.e.\ the magnitude of the star relative to its
particular set of reference stars. By {\it reference level} we mean
the time series of the reference magitude for the L dwarf target,
i.e.\ the magnitude formed from the average of the fluxes of the
target's reference stars (see Bailer-Jones \& Mundt \cite{bjm01}
section 3.3).

\section{Light curve analysis}\label{lightcurve}

The light curves for the target star and reference stars were plotted
and examined visually for features. The \chisq\ test as used by
Bailer-Jones \& Mundt \cite{bjm01} was also applied to look for
variability across the whole data set and within individual
nights. When interpreted as probabilities of variability on the
assumption of Gaussian errors, these \chisq\ values often indicated
significant variability ($p<<0.01$) in both the target and some of the
reference stars. Assuming that not all of these really are
intrinsically variable at this level, this implies one or more of the
following: 1.\ the photometric errors have been underestimated; 2.\
the errors are non-Gaussian; 3.\ the {\it relative} photometry is not
representative of the {\it intrinsic} brightness of the stars. Point
(1) seems unlikely given the thorough testing of the method described
in section~\ref{photometry} to evaluate the errors. Point (2) is
generally true due to outliers: this will inflate all \chisq\ values
for all stars, meaning that one should adopt a conservative threshold
for flagging variability (i.e.\ a small value of the probability, $p$,
or equivalently a higher \chisq\ value). However, this would not be
magnitude dependent so would not lead to a differential effect between
the target and reference stars. Thus the relative values of \chisq\
can still be used as an indication of variability.  Point (3) could be
problematic on account of colour effects in the extinction variation,
as will be discussed in section~\ref{discussion}.

As we have near-simultaneous observations in two bands, J and \km, we
introduce a parameter to look for correlated changes in the relative
magnitudes. For a given star at epoch number $t$, this parameter is
defined as
\[
Q(t) = \sum_{t'=1}^{t'=t} \frac{m_{\rm J}(t')-m_{\rm J}(t'-1)}{\delta m_{\rm J}}
		. \frac{m_{\rm{K}'}(t')-m_{\rm{K}'}(t'-1)}{\delta m_{\rm{K}'}}
\]
where $m_{\rm J}(t')$ is the relative J magnitude of the star at epoch
$t'$ and similarly for $m_{\rm{K}'}(t')$. $\delta m_{\rm J}$ is the
error in $m_{\rm J}(t')-m_{\rm J}(t'-1)$, obtained from the quadrature
sum of the empirical error measures, $\delta m$, for each epoch (see
section \ref{photometry}). Each of the two terms on the right hand
side of this equation gives the change in relative magnitude
from one epoch to the next in the units of the random error;
significant changes have a modulus larger than unity. As the two
terms, one for J and one for \km, are multiplied together, correlated
changes in the two bands give a positive contribution to the sum;
anticorrelated changes a negative contribution. Thus if the star
shows a series of correlated changes, $Q$ will become more positive;
if they are anticorrelated it becomes more negative. If changes are
random, i.e.\ sometimes correlated and sometimes anticorrelated, $Q$
will do a random walk; in particular it has an expectation value of
zero and variance of $t$ (the number of epochs).  By comparing the
variation in $Q$ for the target with that for its reference stars, we
can see whether the target tends to show a greater level of colour
correlated (or colour anticorrelated) changes than the reference
stars.  This is summarized by $Q_{\rm s}$, the difference between the
maximum and minimum values of $Q$ across all epochs.

As cool spots and dust features effect the radiative flux over a wide
wavelength range in the optical and infrared, the evolution of such
features will produce {\it some} kind of correlated or
anticorrelated change in the J and \km\ bands and hence some pattern
in the Q parameter. This is the motivation
for the Q parameter.  The specific relative amplitudes in these bands
depends on the physical mechanism (spots or clouds or something else)
and of course on the specific atmospheric models used. For example,
making predictions based on the models by Allard et al.\
\cite{allard01}, we see that dust variations in early L dwarfs produce
anticorrelated J/\km\ variations, whereas star spots produce
correlated ones (Bailer-Jones \cite{bj02}). Different relative
amplitudes may be predicted when making different assumptions in the
models, e.g.\ with a different dust grain size distribution or
treatment of convection.

Our J and \km\ observations are not strictly simultaneous. However,
for SDSS1203 and 2M1439, a strict pairing of \km\ observations
immediately after J observations was maintained in the final light
curves, and $Q$ has been evaluated using this pairing.  For various
reasons this was not possible with 2M1112, so $Q$ values have not been
calculated for this field. A summary of the relative photometry,
errors and $Q_{\rm s}$ is given in Table~\ref{relphot}.

\begin{table}
\caption[]{Photometry of targets and their reference stars.  J$_{\rm
r}$ and \km$_{\rm r}$ are the magnitudes of the reference stars and
(J-\km)$_{\rm r}$ their colours, relative to the target (and averaged
across all epochs). $\overline{\delta m_d}$ is the average across
epochs of the relative magnitude errors, as given by equation 4 in
Bailer-Jones \& Mundt \cite{bjm01}. $Q_{\rm s}$ is defined in the
text.}
\label{relphot}
\begin{tabular}{lrrrrrr}
\hline
         & J$_{\rm r}$ & $\overline{\delta m_d}$ & \km$_{\rm r}$ & $\overline{\delta m_d}$ & (J-\km)$_{\rm r}$ & $Q_{\rm s}$ \\
\hline
SDSS1203 &  0.000 & 0.016 &  0.000 & 0.012 &  0.000 &  9 \\
ref.\ 1  & -1.620 & 0.010 & -0.849 & 0.013 & -0.771 &  7 \\
ref.\ 2  &  0.578 & 0.012 &  0.448 & 0.014 & -1.026 & 15 \\
ref.\ 3  &  0.829 & 0.021 &  1.757 & 0.035 & -0.928 & 11 \\
ref.\ 4  &  0.640 & 0.019 &  1.614 & 0.030 & -0.974 &  7 \\
ref.\ 5  &  0.932 & 0.026 &  1.701 & 0.033 & -0.769 &  9 \\
ref.\ 6  &  1.467 & 0.037 &  2.562 & 0.067 & -1.095 &  7 \\
ref.\ 7  &  1.242 & 0.035 &  1.516 & 0.032 & -0.274 &  6 \\
ref.\ 8  &  1.433 & 0.033 &  2.376 & 0.053 & -0.943 & 10 \\
\hline
2M1439   &  0.000 & 0.016 & 0.000 & 0.018 &  0.000 & 15 \\
ref.\ 1  & -0.559 & 0.017 & 0.101 & 0.023 & -0.660 &  8 \\
ref.\ 2  &  0.707 & 0.018 & 1.431 & 0.024 & -0.724 & 23 \\
ref.\ 3  &  1.953 & 0.026 & 2.699 & 0.042 & -0.746 &  8 \\
ref.\ 4  &  2.257 & 0.034 & 3.102 & 0.058 & -0.845 & 12 \\
ref.\ 5  &  2.557 & 0.038 & 2.901 & 0.050 & -0.344 & 15 \\
\hline
\end{tabular}
\end{table}

\section{Results}\label{results}

\begin{figure*}
\begin{minipage}{150mm}
\centerline{
\includegraphics[width=\textwidth]{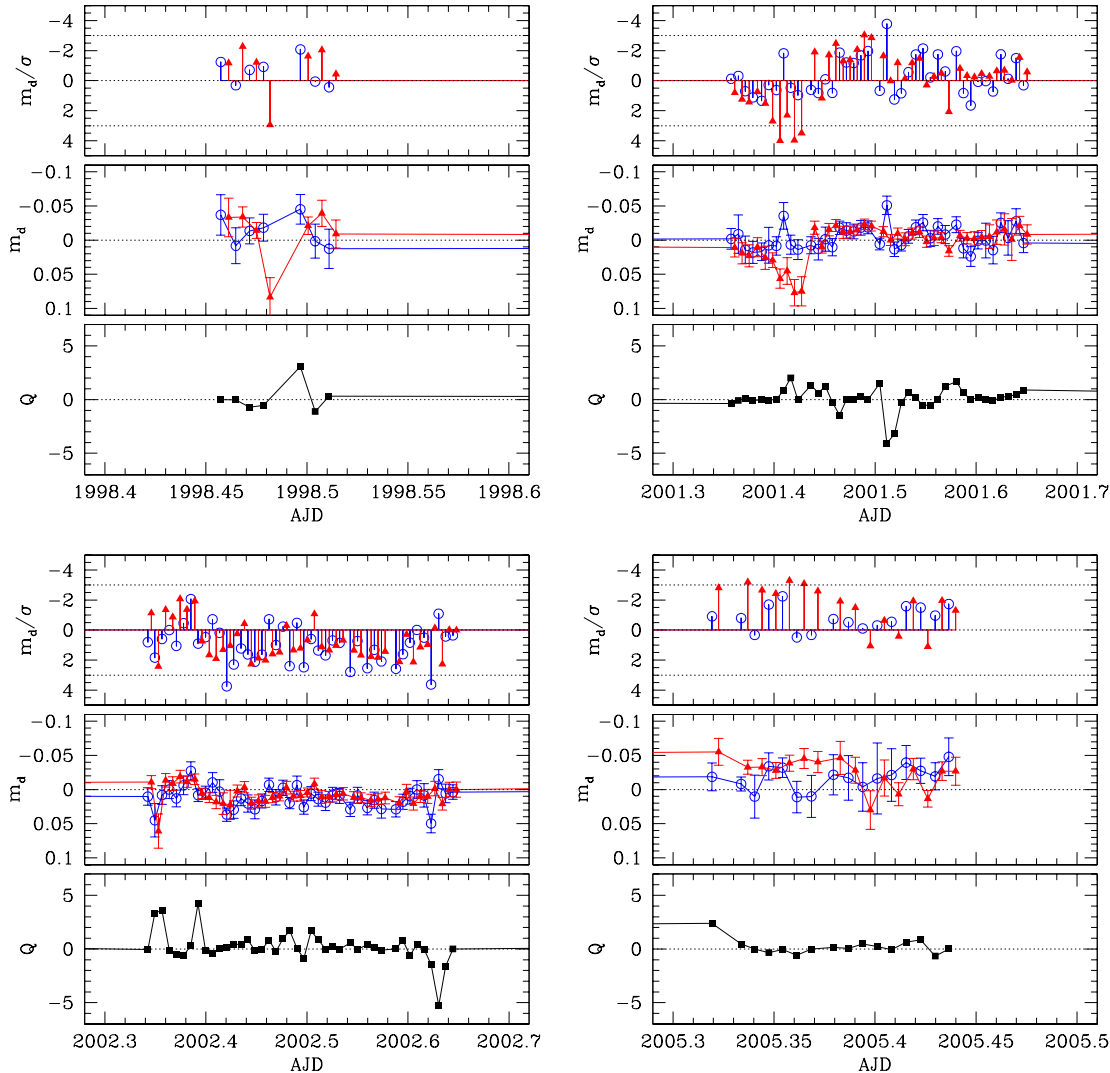}
}
\caption{Light curves for SDSS1203, one plot for each night (nights 10, 13,
14 and 17). The open circles (blue) are for the J band and the closed
triangles (red) for the \km\ band. The middle panel for each night shows
the relative light curve of SDSS1203 (with a common magnitude zero
point for all nights equal to the average relative magnitude across all
epochs). The top panel for each night shows this light curve in units of
the photometric errors at each epoch, which can be used to judge the
significance of any deviations. The bottom panel plots the $Q(t)$
value, defined by the equation in section \ref{lightcurve}.}
\label{sdss1203_all}
\end{minipage}
\end{figure*}

\noindent{\bf SDSS1203} Calculated across the entire data set,
SDSS1203 gives \chisq\ values of 200 and 306 for J and \km\
respectively.  These formally correspond to very small $p$ values (a
test with 103 degrees of freedom) but, as noted in section
\ref{lightcurve}, we may expect \chisq\ to be large due to outliers
across this large number of observations. Indeed, the reference stars
also give large \chisq\ values: ranging from 114 to 158 (for J) and 98 to 155 (for
\km).  Ignoring the formal $p$ values and just looking at the relative
\chisq\ values, we  see that SDSS1203 is indeed more variable than any of
the reference stars in both filters.

Fig.\ \ref{sdss1203_all} shows the J and \km\ light curves across the
four nights of observation. There are several points which deviate
from the mean by more than 3$\sigma$, although generally the
variability is quite weak ($<0.05$ mag deviation).
On night 13, there is a dip in \km\ at around AJD 2001.4.  A similar
but opposite effect is seen in reference star 2 (although not ref.\ 1)
and the egress from this dip is accompanied by a rise in the reference
level, so this is probably a telluric effect.
The 4$\sigma$ jump in J at AJD 2001.51 is not accompanied
by any similar feature in the reference stars or reference level so
could be intrinsic to SDSS1203, although it is neither large nor that
statistically significant. The J dip at AJD 2002.63, on the other
hand, {\it is} accompanied by a dip in the reference level (but not
the reference stars). This could be due to a colour dependent
variation in the extinction (see section \ref{discussion}).
We also note that SDSS1203 appeared to be brighter in \km\ at the
beginning of night 17, an effect which is marginally significant and
cannot obviously be reduced to the behaviour of the reference stars or
reference level.

The $Q$ parameter in Fig.\ \ref{sdss1203_all} shows no long lasting
colour correlated variations.  Moreover, the range of $Q$, given by
$Q_{\rm s}$, is no larger than for the reference stars (Table
\ref{relphot}). The J/\km\ correlation coefficient, $\rho$, is
slightly larger for SDSS1203 than the other stars ($\rho$=0.34 against
$\rho$=0.07--0.24 for the reference stars), although it is still
small.

We conclude that there is no good evidence for variability in SDSS1203
from these data.  SDSS1203 was found to be variable in the I-band by
Bailer-Jones \& Mundt \cite{bjm01}, primarily due to an apparent
brightening of the object over a duration of one to two hours.  As far
as we could tell, those data were taken under good observing
conditions near culmination of the object (i.e.\
very small airmass changes) so we do not believe that result was an
artifact due to the Earth's atmosphere.  ~\newline

\begin{figure*}
\begin{minipage}{150mm}
\centerline{
\includegraphics[width=\textwidth]{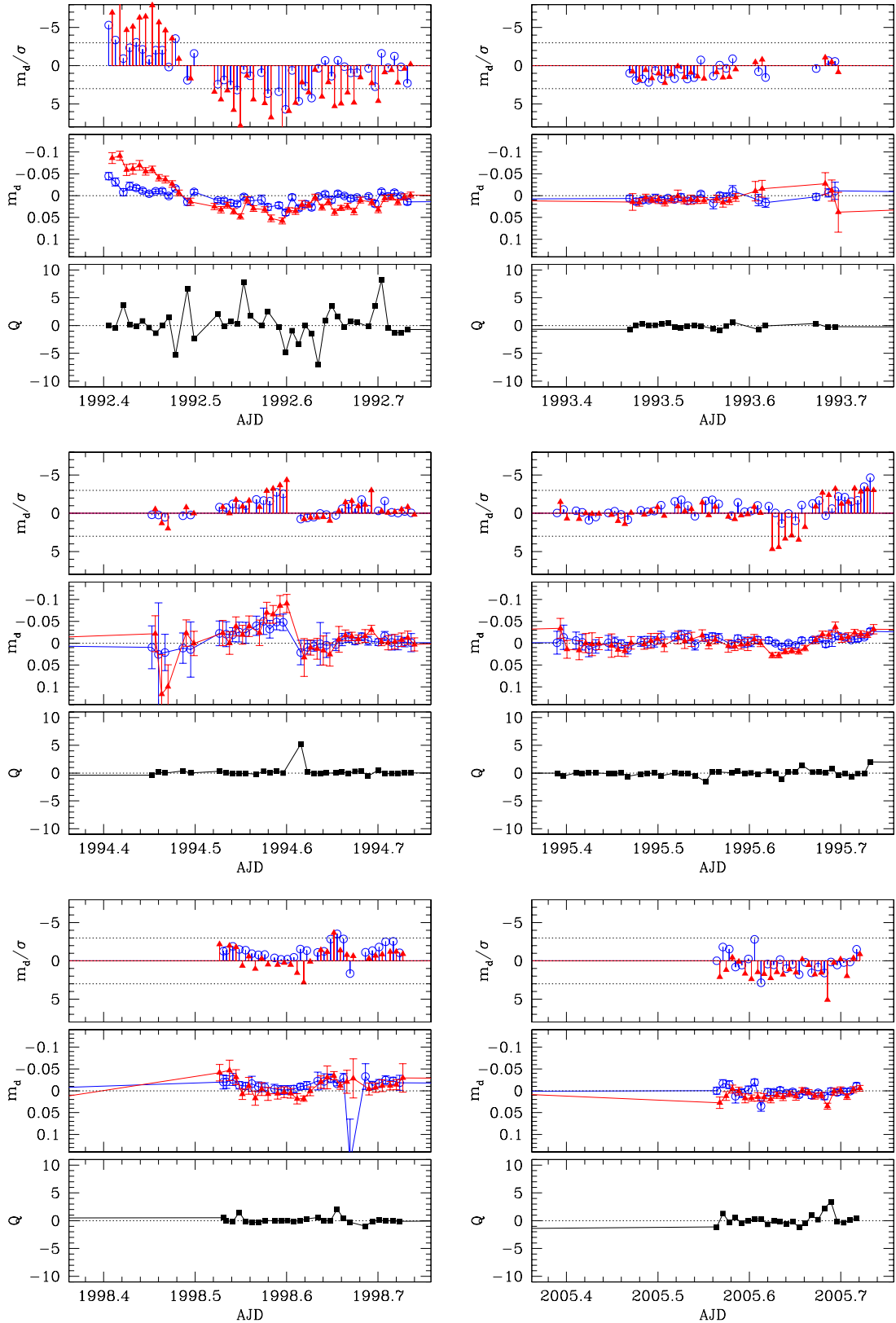}
}
\caption{Light curves for 2M1439 (see caption to Fig.\
\ref{sdss1203_all} for an explanation). Light curves are shown only
for 6 of the 11 nights on which data were taken (see Table
\ref{obslog}), specifically for nights 4, 5, 6, 7, 10 and 12.}
\label{2m1439_all}
\end{minipage}
\end{figure*}

\begin{figure*}
\begin{minipage}{180mm}
\centerline{
\includegraphics[width=0.83\textwidth]{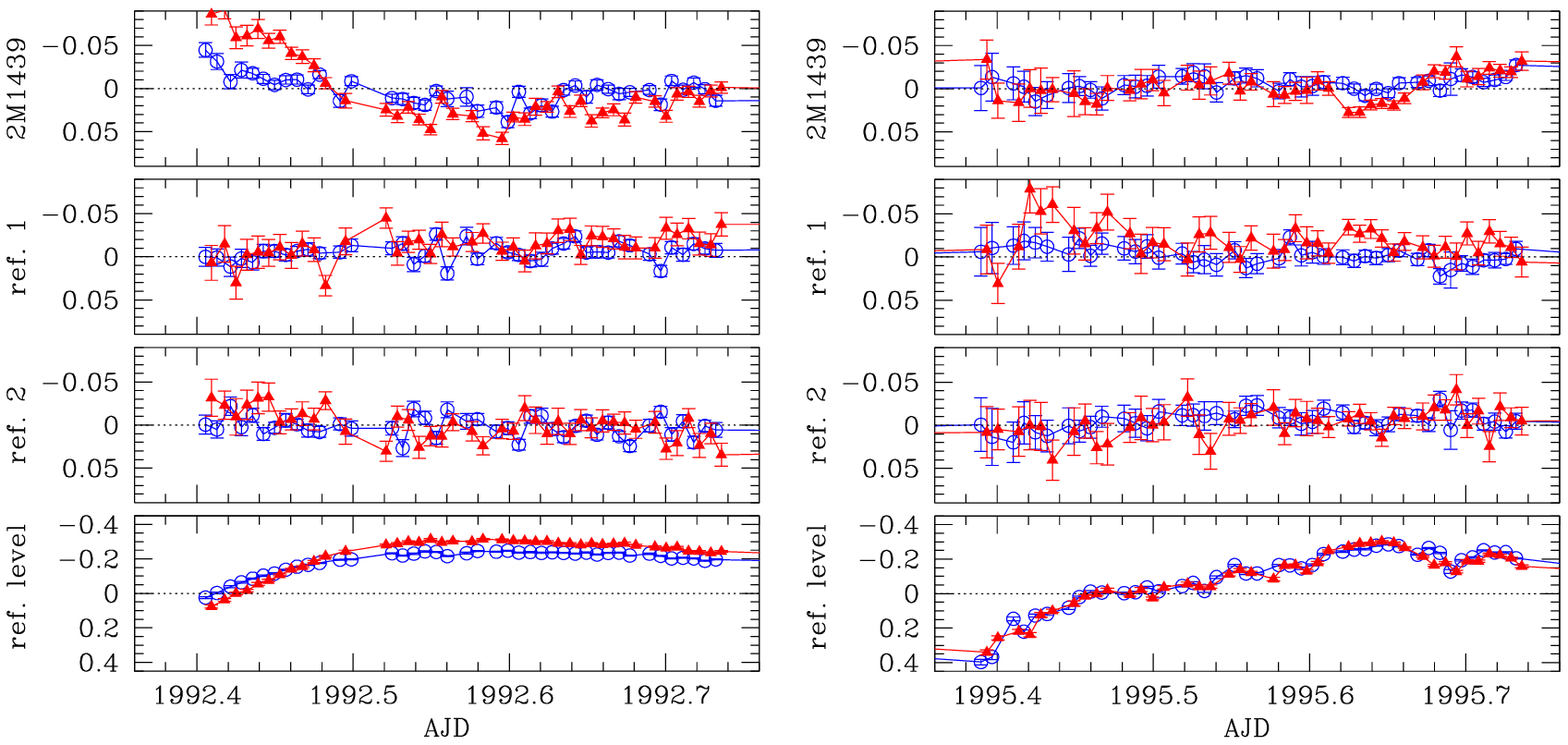}
}
\caption{Relative magnitude light curves for 2M1439 (top row), the
first (i.e.\ brightest) reference star (second row) and the second
(i.e.\ next brightest) reference star (third row). The bottom row
shows the variation in the reference level for 2M1439. The reference
level is the magnitude formed from the average of the fluxes of all
five reference stars shown in Table \ref{relphot} (with the average of
this level over all epochs subtracted). As in the other figures, the
open circles (blue) are for the J band and the closed triangles (red)
for the \km\ band.  The left panel is for night 4, the right panel for
night 7.}
\label{2m1439multi_lc}
\end{minipage}
\end{figure*}

\noindent{\bf 2M1439} As with SDSS1203, this object shows more
significant variability than any of the reference stars for both J and
\km\, according to the \chisq\ test applied to all data. However, its
$Q$ values, as well as the correlations between J and \km\ and J and
J$-$\km\ are no larger than for some of the reference stars.

It can be seen in Fig.\ \ref{2m1439_all} that the \km\ light curve of
2M1439 dims by 0.14 mag over a period of five hours from AJD 1992.4 on
night 4.  Although no such trend (or opposite trend) is seen in the
light curve of the reference stars individually, the opposite trend is
seen in the reference level with a scale of 0.4 magnitudes, indicating
that all of the reference stars increased in brightness (Fig.\
\ref{2m1439multi_lc}). This coincides with the field rising -- from an
airmass of 2.6 to 1.05 -- so is presumably just due to the usual
decrease in extinction with decreasing airmass. However, while the
reference stars brighten by 0.4 mag, 2M1439 brightens by only 0.26
mag. This is possible if the extinction coefficient has a strong
gradient across the \km\ band and if 2M1439 has a different flux
gradient from the reference stars.  In this case, second order
(colour) effects in the terrestrial extinction may become important
(see section~\ref{discussion}).  The same effect is seen in the J band
(reference level brightens by 0.35 mag, 2M1439 dims relative to this
by 0.07 mag).

However, on night 7 we see the same brightening of the reference level
due to decreasing airmass (0.65 mag in both bands from airmass 2.8 to
1.05) but without any trend or significant variability in the light
curves of either the reference stars or 2M1439 (see right hand panel
in Fig.\ \ref{2m1439multi_lc}). Clearly, different atmospheric conditions
must have prevailed on this night than on night 4.

Correlated changes of up to 1 magnitude in J and \km\ are found in the
reference level at the beginning of night 6, caused by thin clouds and
humidity.  Correlated changes with amplitude up to 0.1 mag are also
seen in the light curves of the reference stars and 2M1439, indicating
that these large extinction variations have not cancelled out in the
relative photometry.  This is presumably caused by water vapour
changes associated with the clouds (see section \ref{discussion}).

In conclusion, most of the features which account for the higher
variability in 2M1439 (compared to the reference stars) appear to be
related to terrestrial atmospheric phenomena.

2M1439 has previously been monitored in the I-band by Bailer-Jones \&
Mundt \cite{bjm01} and by Gelino et al.\ \cite{gelino02}. Neither
found good evidence for variability above about 0.01 mag, the former
from temporally-dense observations (48 observations spread over
four nights) and the latter over longer timescales (a few observations
per night over 40 nights).  ~\newline

\noindent{\bf 2M1112} This light curve shows some variations which at
first sight would be of significance, but on closer inspection appear
to be due to variations in the reference level.  These are not
qualitatively different from the ``problematic'' variations seen in
the other two targets, discussed above.
~\newline

\noindent We performed a frequency-domain search for periodic
variations using two methods, the CLEAN and Lomb-Scargle periodograms.
These methods yielded consistent results and showed no evidence for
periodic variations in SDSS1203 or 2M1439. As implemented, these
methods are known to be effective at detecting even low
signal-to-noise periodic signals (Lamm et al.\ \cite{lamm03}).

\section{Second order extinction effects}\label{discussion}

The method of relative photometry employed in this program (and all
those referred to in section \ref{introduction}) makes the implicit
assumption that the Earth's atmospheric extinction is the same for the
target star as for the reference stars.  However, the extinction
coefficient varies with wavelength, so if the target and reference
stars have different spectral energy distributions (SEDs), their
wavelength integrated extinctions are also different. This is relevant
when monitoring ultra cool dwarfs, as the reference stars will usually
be significantly hotter and hence bluer in the optical and near infrared
(see Table ~\ref{relphot}).  This would not matter for relative
photometry if the extinction did not vary in time or space. But 
extinction does vary polychromatically with airmass and atmospheric conditions, 
thus introducing a change in the relative magnitude
between stars with different SEDs. Such effects are referred to as
second order extinction.

In the visible, the extinction is dominated by molecular (Rayleigh)
and aerosol scattering and generally shows a smooth decrease with
increasing wavelength. In this case, colour effects are either small
or a broad band colour term can be used to correct relative photometry
(also as a function of airmass). In the near infrared, however, the
extinction is dominated by molecular absorption, particularly of \h2o,
\co2\ and \o3, which show much sharper variations with
wavelength. Precipitable water vapour and ozone concentrations in particular can
vary rapidly in time and space. 
{\it Thus even under apparently good observing conditions, second
order extinction effects can have a significant impact on high
precision, differential, broad band infrared photometry.}  Note that
terrestial water clouds and fog are dominated by large
($>>$\,1\,\micron) water droplets, so these cause a scattering of
visible and infrared light with only a weak wavelength dependence
($\lambda^{- \alpha}$, where $0< \alpha <1$).  Nonetheless, some
absorption also occurs in the liquid drops themselves which does show
rapid changes in some wavelength ranges, although in the optical at
least the colour dependence of terrestrial cloud extinction appears to
be small (Honeycutt \cite{honeycutt71}; Serkowski
\cite{serkowski70}). However, where there are clouds there is
presumably also a high water {\it vapour} column density, with the result
that clouds would contribute to second order extinction.

Unfortuntately, we cannot quantify the effects of second order
extinction on our data, because we have no independent measure of the
atmospheric constituents -- in particular the column density of the
precipitable water vapour -- and we would further require an appropriate
atmospheric model matched to the Calar Alto conditions at the
time. Moreover we do not know the SEDs of the reference stars: the
broad band J$-$\km\ colour is too undersampled to determine the
integrated extinction from a water spectrum (Young
\cite{young89}). 

An idea of the scale of the problem can be found from reference to the
models of J and K extinction at Kitt Peak for a cool giant
(\teff=4000\,K) and Vega (\teff=9650\,K) from Manduca \& Bell
(1979). At low water column density, the differential extinction
between the two stars (E$_{\rm 4000}-$E$_{\rm 9650}$) is 0.049 mag at
J and 0.002 mag at K.  Increasing the water column density by a factor
of 35 changes these to 0.067 mag at J and 0.001 mag at K.  Thus
increasing the precipitable water density increases the differential
extinction (so decreases the apparent brightness of a cool star
relative to a hotter star) by approximately 0.02 mag at J and $-$0.001
mag at K.\footnote{Derived from Tables IIa and IIb of Manduca \& Bell
for Johnson filters at an airmass of 1.5.}  The actual precipitable
water column density experienced during our observations could be
larger, so it is not implausible that water vapour changes cause the
apparent variability we see in the L dwarfs.  Contrary to the figures
given above, however, we generally see similar amplitude variations in
J as in \km. But our \km\ filter extends further into the
blueward telluric absorption band than does the Johnson K band of
Manduca \& Bell and these figures are quite sensitive to the band
profile. Also, our target (and probably reference) stars have much
lower temperatures than 4000\,K and 9650\,K.

Changes in airmass under stable conditions can also change the
relative photometry.  For example, Manduca \& Bell show that the J
band magnitude of a cool giant relative to Vega would be at least 0.01
mag higher when measured at airmass 2.0 than at airmass 1.0.  The
effect is much smaller in the K band, however, so this may not be
cause of the dimming of 2M1439 on night 4 discussed in section
\ref{results} (see Fig. \ref{2m1439multi_lc}), as there the dimming is
larger in \km\ than J. However, this may again be due to the
differences between the K and \km\ bands and the SEDs of the stars.
Recall that no such dimming was seen on night 7 for the same airmass
change, perhaps indicating different causes of extinction on these
nights.

The fact that both SDSS1203 and 2M1439 showed larger \chisq\ values
for J and \km\ than the reference stars can be attributed to the
likelihood that the reference stars have more similar SEDs to each
other than to L dwarfs, so that differential extinction effects
between any one reference star and all the other reference stars (which form
its reference level) are much smaller. Hence the reference stars show
less variability.

Second order extinction is less of a problem in the optical
($\lambda<0.65$\,\micron), but here most brown dwarfs are far too
faint to be monitored. The I band ($\sim$\,0.78--0.92\,\micron) is a
good compromise. This too is intersected by a water absorption band
centered at 0.81\,\micron, but this band is much weaker than those
effecting the J and K bands (e.g.\ McCord \& Clark \cite{mccord79}).

\section{Conclusions}

The prime requirements of
any search for variability are (1) a measurement of the scale of the
variability {\it intrinsic} to the source, and (2) an accurate
determination of the observational errors in these measurements.
However the analysis is done, a variability detection relies on the
former being significantly larger than the latter.

We have developed an observation and data reduction technique which 
gives an accurate and reliable determination of the
photometric errors, arising from the source, background, detector and
effect of the data processing (see
section~\ref{photometry}).

One of the main problems with ground-based monitoring is variations in
the extinction (scattering plus absorption) of the Earth's atmosphere.
The standard procedure of differential photometry was used
to give a measure of intrinsic variability.  Using this, some evidence
for variability was found in the observed L dwarfs.  However, closer
analysis has shown that this measure is probably ``contaminated'' by
second order extinction effects in the Earth's atmosphere
(see section \ref{discussion}).  Although we have no direct evidence
for this, it seems the most plausible cause of the observed
variability given the features we have described and the elimination
of other potential causes.  We therefore conclude that we have no good evidence
for intrinsic variability in any of the three L dwarfs monitored.
None of these objects showed significant correlations in their J and
\km\ light curves, as evidenced by correlation plots and our $Q$
parameter. Some such (anti)correlation would be expected if the variability
is caused by cool spots or dust features.  Upper limits on the
peak-to-peak amplitude of {\it persistent} variability are set by the
scatter in the light curves under the most stable Earth atmospheric
conditions. For both J and \km, these limits are set at 0.04 mag for
SDSS1203 and 2M1439 and 0.08 for 2M1112.

As late M, L and T dwarfs are at their brightest in the near infrared
(0.9--2.5\,\micron) and show spectral signatures which could
discriminate between various surface feature scenarios (Bailer-Jones
\cite{bj02}), this is a desirable wavelength range for variability
monitoring.  We have shown that high {\it precision} differential
photometry (random errors of less than two percent) can be
achieved on L dwarfs with infrared arrays.  However, as a measure of
the intrinsic variability in L dwarfs, broad band differential
infrared photometry appears to be limited in {\it accuracy} to a few
percent by variations in terrestrial molecular extinction, in
particular precipitable water vapour. These second order extinction
effects could be reduced in one or more ways:
\begin{enumerate}
\item{use of passbands which avoid molecular extinction, in particular
use of better designed (narrower) J and \km\ filters to avoid the
strong \h2o\ bands which generally intersect these ``standard''
filters\footnote{There is significant variance between the profiles of
  filters called I, Z, J, H and K at different observatories, so there
  is no such thing as a ``standard'' infrared filter set, although most Z, J
  and K filters are intersected by significant water absorption bands.};}
\item{use of narrower band filters, which reduces the effect of second order
extinction\footnote{A variation on this method has been used by
  Bailer-Jones \cite{bj02}, in which an L dwarf was monitored
  spectrophotometrically relative to another star in the spectrograph
  slit. The relative photometry was then performed independently in
  each narrow wavelength bin (0.002 \micron), reducing effects of
  second order extinction to well below other error sources.} (this would
obviously require a larger telescope or longer integration times; the
latter will be unacceptable for detecting short-term
variability);}
\item{use of specially designed filters to monitor the dominant
molecular absorption (in particular \h2o)\ from which the broad band
relative photometry could, in principle, be corrected (Angione
\cite{angione87}). This could be done using a multi-channel
camera. However, additional spectroscopic observations are required to
calibrate the measures and so apply the correction. Nonetheless, some
kind of time-dependent monitoring of extinction would at least allow
one to recognise when observing conditions were variable;}
\item{observing from very dry sites under atmospherically stable
conditions.
Space-based observations remove all problems related to
second order extinction. A strong discriminant between
candidate surface features in brown dwarfs is the variability
signatures in the water absorption bands at 1.35--1.45\micron\ and
1.80--2.00\micron\ (e.g.\ Bailer-Jones \cite{bj02}. These bands cannot be
reliably monitored from the ground, providing another argument for
space-based observations.}
\end{enumerate}

\section*{Acknowledgements}

We are grateful to Ken Mighell for use of his CCDCAP algorithm and for
useful information on its application and interpretation.

\end{document}